\documentclass[doublecol]{epl2}
\usepackage{amsmath,amssymb,graphicx}
\usepackage{bm}
\definecolor{labelkey}{cmyk}{.4,.2,0,0}
\renewcommand{\epsilon}{\varepsilon}
\usepackage{graphicx,color}
\graphicspath{{./}{./figures/}{./Figures/}}
\usepackage{dcolumn}
\usepackage{bm}

\newcommand{\Eq}[1]{Eq.~(\ref{#1})}
\newcommand{\eq}[1]{(\ref{#1})}
\newcommand{\half}{\frac12}

\newcommand{\bea}{\begin{eqnarray}}
\newcommand{\eea}{\end{eqnarray}}
\newcommand{\beq}{\begin{equation}}
\newcommand{\eeq}{\end{equation}}

\newcommand{\rme}{\mathrm{e}}
\newcommand{\rmd}{\mathrm{d}}
\newcommand{\nn}{\nonumber}
\newcommand{\nott}[1]{}
\newcommand{\Fig}[1]{\includegraphics[width=\columnwidth]{#1}} 
\newcommand{\fig}[2]{\includegraphics[width=#1\columnwidth]{#2}}

\graphicspath{{figures/},{}}

\renewcommand{\paragraph}[1]{{\it #1}}

\begin{document}

\title{Avalanche shape and exponents beyond mean-field theory}
\author{Alexander Dobrinevski, Pierre Le Doussal, Kay J\"org Wiese}
  \institute{CNRS-Laboratoire de Physique Th\'eorique de l'Ecole Normale
  Sup\'erieure, 24 rue Lhomond, 75005 Paris, France.}

\abstract{Elastic systems, such as magnetic domain walls, density waves, contact lines, and cracks,
are all pinned by substrate disorder. When driven, they move via successive jumps 
called avalanches, with power law distributions of size, duration and velocity.
Their exponents, and the shape of an avalanche, defined as
its mean  velocity  as  function of time, have recently been studied.
They are known approximatively from experiments and simulations, and 
were predicted from mean-field models, such as the Brownian force model (BFM), 
where each point of the elastic interface sees a force field which itself is a random walk.
As we showed in EPL {\bf 97} (2012)   46004,    the BFM is the starting point for an $\epsilon = d_{\rm c}-d$ expansion 
around the upper critical dimension, with $d_{\rm c}=4$ for short-ranged  elasticity, and $d_{\rm c}=2$ for long-ranged  elasticity. 
Here we  calculate analytically the ${\cal O}(\epsilon)$, i.e.\ 1-loop, correction to the {\it avalanche shape at fixed duration} $T$,
for both types of elasticity. The exact expression is well approximated by  
$\left< \dot u(t=x T)\right>_T\simeq  [ Tx(1-x)]^{\gamma-1} \exp\left( {\cal A}\left[\frac12-x\right]\right)$, $0<x<1$.    
The asymmetry ${\cal A}\approx - 0.336 (1-d/d_{\rm c})$ is negative for $d$ close to $d_{\rm c}$, 
skewing the avalanche towards its end, as observed in numerical simulations in $d=2$ and $3$. 
The exponent $\gamma=(d+\zeta)/z$ is given by the two independent exponents at depinning, the roughness $\zeta$ and the dynamical exponent $z$.
We propose a general procedure to predict other avalanche exponents in terms of $\zeta$ and $z$. 
We finally introduce and calculate
the shape {\it at fixed avalanche size}, not yet measured in experiments or simulations.}

\pacs{68.35.Rh}{Phase transitions and critical phenomena}

\maketitle

\paragraph{Introduction:}
An elastic interface driven through a disordered medium is an efficient mesoscopic model for a number of different physical systems, such as the motion of  domain walls in soft magnets  \cite{DurinZapperi2000}, fluid contact lines on a rough surface \cite{LeDoussalWieseMoulinetRolley2009}, or strike-slip faults in geophysics; see \cite{DSFisher1998} for a review.
Their response to external driving is not smooth, but exhibits collective jumps called \textit{avalanches}, extending over a broad range of space and time scales. They can be detected e.g.~as pulses of Barkhausen noise in magnets \cite{Barkhausen1919,DurinZapperi2006b}, slip instabilities leading to earthquakes on geological faults, or in fracture experiments \cite{BonamyPonsonPradesBouchaudGuillot2006}. While the microscopic details of the dynamics are specific to each system,
an important question is whether the large-scale features are universal. A  prominent example are the exponents 
of the power-law distribution function (PDF) of avalanche sizes $P(S) \sim S^{-\tau}$ (for earthquakes, the  Gutenberg-Richter law) and durations. Beyond scaling exponents,
the question of whether the shape of an avalanche is universal is of great current interest in theory and experiments~\cite{PapanikolaouBohnSommerDurinZapperiSethna2011,LaursonIllaSantucciTallakstadyAlava2013}.
Understanding how universality arises,  which quantities are universal, and how to make quantitative predictions beyond phenomenological models
are some of the main challenges in the field. 

Historically, the elastic-interface model allowed for analytical progress 
thanks to a powerful method, the Functional Renormalization group (FRG). It was first developed to calculate the static (equilibrium) deformations
of an interface pinned by a random potential (e.g.\ the roughness exponent),
or the critical dynamics at and beyond the depinning transition,  applying an external force $f \ge f_{\rm c}$
\cite{DSFisher1986,NattermannStepanowTangLeschhorn1992,NarayanDSFisher1992b,ChauveLeDoussalWiese2000a}. These results were obtained in an expansion in the 
internal space dimension $d$ of the interface around the upper critical dimension $d_{\rm c}$, equivalent to 
 a loop expansion. Despite these successes, the study of avalanches in elastic systems has remained
centered on toy models \cite{DSFisher1998,ABBM,Colaiori2008}, scaling arguments, and numerics 
\cite{MiddletonFisher1993,NarayanMiddleton1994,LuebeckUsadel1997,NarayanDSFisher1992b,ZapperiCizeauDurinStanley1998}.
Other important models  used to
describe avalanches are  the random-field Ising model \cite{BanerjeeSantraBose1995+}, 
 mean-field spin glasses \cite{avalspinglasses},
and discrete automata alias sandpile models, with some 
analytical results \cite{BakTangWiesenfeld1987,IvashkevichPriezzhev1998,Dhar1999}. 
However, exact results on the avalanche statistics are notably hard to obtain. 
Recently, we have extended the FRG-based field theory to calculate the avalanche-size distribution \cite{LeDoussalMiddletonWiese2008,LeDoussalWiese2008c} in dimension $d<d_{\rm c}$, with excellent agreement to numerics \cite{LeDoussalMiddletonWiese2008,RossoLeDoussalWiese2009a}. We then extended the theory to the dynamics and obtained the velocity distribution within an avalanche \cite{LeDoussalWiese2011a}. 

In this Letter, we use this theory to propose several novel scaling relations for avalanche exponents, and calculate the shape of an avalanche,
both at fixed duration and at fixed size.  
Since the calculations are very technical, we only sketch the main ingredients of the method and present the key results; the details are given  in a separate publication \cite{LeDoussalWieseToBePublished}. For an early presentation of this work 
see  \cite{DobrinevskiPhD}.

\medskip

\paragraph{Avalanche densities and dynamical action:}
Consider the equation of motion for a driven elastic interface in presence of quenched disorder\footnote{We use indifferently $\partial_t$ or a dot for time derivatives.},
\beq
\eta \partial_t u_{xt} = \nabla_x^2u_{xt}   - m^2 u_{xt} + f_t+ F(u_{xt},x)\ . \label{eqmo1} 
\eeq
We denote by subscript the dependence on space and time. Choosing $f_t=m^2 w_t$ the interface is bound by a parabolic well of curvature $m^2$ (the mass) to an external degree of freedom $w_t$. The pinning force $F(u,x)$ is chosen Gaussian with (microscopic) correlator (overlines denote disorder
averages)
\beq  \label{BFM-correl}
\overline{ F(u,x) F(u',x')} = \delta^d(x-x') \Delta_0(u-u').
\eeq
Intermittent avalanche motion occurs for slow driving, either at small constant
velocity $w_t=v t$, or upon a small force step, i.e.\ a kick $\dot w_t=w \delta(t)$.
Avalanche-size and duration distributions, as well as the shape, can be retrieved from the
generating function, i.e.\ the disorder average of
\beq\label{9}
G[\lambda,f]:= \overline{ \rme^{\int_{x,t} \lambda_{xt} \dot u_{xt}} }^f ,
\eeq
in presence of a source $\lambda_{xt}$ and  a driving force $f_{t}=m^{2} w_{t}$.
For instance, the PDF of the size of an avalanche, $S:=\int_{x,t>0} \dot u_{xt}$, following a kick $f_{t}=m^2 w \delta(t)$,
is the inverse Laplace transform, $P_w(S)= \mbox{LT}^{-1}_{-\lambda \to S} G[\lambda,f]$ 
for a uniform source $\lambda_{xt}=\lambda$. From it
one defines a size density (per unit displacement $w$),
$\rho(S):=\partial_w P_w(S)|_{w=0^+}$, which equals the size density defined from stationary motion\footnote{We take advantage of the {\em Middleton
theorem} \cite{Middleton1992} which ensures forward-only motion for forward driving,
and prepare the system in the {\em Middleton attractor}, as discussed in  Refs.~\cite{DobrinevskiLeDoussalWiese2011b,LeDoussalWiese2011a}.} at fixed $v=0^+$. Similarly
 one defines  the density for the avalanche duration $T$. 
All these densities, for sizes $S \ll S_m$ and times $T \ll \tau_m$, obey 
power laws with exponents
\beq \label{densities} 
 \rho(S) \sim S^{-\tau} \quad , \quad \rho(T) \sim T^{-\alpha} \ ,
\eeq 
where $S_m$ and  $\tau_m$  are set by the mass (and $\eta$ for $\tau_{m}$) and are used as convenient units below\footnote{\label{footnote4}
Both can be {\it measured}, $S_m:=\langle S^2 \rangle/(2 \langle S \rangle)$ 
from the moments of the size PDF \cite{LeDoussalWiese2008c}, and $\tau_m$ from the
response function \cite{LeDoussalWiese2011a,DobrinevskiLeDoussalWiese2013}.}.

\begin{table*}[t]
\scalebox{1}{\begin{tabular}{|c|c|c|c|c|c|}
\hline 
 & $\rho(S) $ & $\rho(S_\phi)$ & $\rho(T)$ & $\rho(\dot u)$ & $\rho(\dot u_\phi)$ \\
\hline 
\hline 
    & $S^{-\tau}$ & $ S_\phi^{-\tau_\phi}$ & $T^{-\alpha}$ & $\dot u^{-\sf a}$ & $\dot u_\phi^{{-\sf a}_\phi}$  \\
\hline 
  short-ranged elasticity (SR)      & $\tau = 2 - \frac2{d+\zeta}$ &$\tau_\phi = 2 - \frac2{d_\phi+\zeta}$ & $\alpha = 1+ \frac{d-2+\zeta}{z}$ & ${\sf a} = 2- \frac2{d+\zeta-z}$ & ${\sf a}_{\phi} = 2- \frac2{d_\phi+\zeta-z}$  \\
\hline 
 long-ranged elasticity (LR)      & $\tau = 2 - \frac1{d+\zeta}$ &$\tau_\phi = 2 - \frac1{d_\phi+\zeta}$ & $\alpha = 1+ \frac{d-1+\zeta}{z}$ & ${\sf a} = 2- \frac1{d+\zeta-z}$ & ${\sf a}_{\phi} = 2- \frac1{d_\phi+\zeta-z}$  \\
\hline 
\end{tabular}}
\caption{Scaling relations}
\label{tab1}
\end{table*}%

To calculate $G[\lambda,f]$, one  takes a time-derivative of Eq.~(\ref{eqmo1}),
\beq
\eta \partial_t \dot u_{xt} = \nabla_x^2 \dot u_{xt}  + f_t - m^2 \dot u_{xt} + \partial_t  F(u_{xt},x),
\eeq
and constructs the dynamical field theory by multiplying this equation with a response-field $\tilde u_{xt}$.
Averaging over disorder leads to the path-integral representation
\beq \label{G} 
G[\lambda,f] = \int {\cal D}[\dot u]  {\cal D}[\tilde u]  \rme^{-{\cal S}_{\lambda,f}[u,\tilde u]}\ .
\eeq
The dynamical action reads 
\begin{align}\label{5}
& {\cal S}_{\lambda,f}[u,\tilde u] = {\cal S}_0[u,\tilde u]+ {\cal S}_{\rm dis}[u,\tilde u] + \int_{xt} \lambda_{xt} \dot u_{xt}    \\ 
&\!\!\!{\cal S}_0[u,\tilde u] =\int_{xt} \tilde u_{xt} \left[ \eta \partial_t \dot u_{xt}-\nabla^2 \dot u_{xt} +m^2 \dot u_{xt} - f_t  \right]    \label{6} \\ 
&\!\!\!{\cal S}_{\rm dis}[u,\tilde u] =-\half \int_{x,t,t'} \tilde u_{xt} \tilde u_{x,t'}\partial_t \partial_{t'} \Delta_0(u_{xt}-u_{xt'})\,. 
\end{align}
Upon coarse-graining, the action becomes the effective action 
${\cal S}_{\rm dis} \to {\cal S}^{\rm eff}_{\rm dis}$, 
with a renormalized disorder correlator $\Delta(u)$, which takes a non-analytic form with a linear cusp at $u=0$, 
\beq\label{Delta-ren}
\Delta_0(u) \to \Delta(u) = \Delta(0) -\sigma |u| - \frac{g}2 u^2+ ...
\eeq
with $\sigma=-\Delta'(0^{+})$, and  $g=-\Delta''(0^+)$.
Hence one can rewrite \cite{LeDoussalWiese2011a}
\bea
{\cal S}^{\rm eff}_{\rm dis}[u,\tilde u] &=& -\sigma \int_{x,t}  \tilde u_{xt}^2 \dot u_{xt} \nn\\
&& - \frac{g}{2} \int_{x,t,t'}  \tilde u_{xt}\dot u_{xt} \, \tilde u_{xt'}\dot u_{xt'}  + ...
\label{8}
\eea
Note that while the disorder  interaction is in general  {\em non-loca}l in time, the first term, proportional to $\sigma$, is {\em local}, since $\frac{\rmd ^2}{\rmd u^2} |u| = 2 \delta(u)$; a simplifying feature to be exploited below. 

\medskip

\paragraph{Mean-field theory: the Brownian force model.} Further suppose that the microscopic force correlator (\ref{Delta-ren}) only contains the term $- \sigma |u|$, realized if for each $x$ the forces $F(u,x)$ are chosen as Brownian motions in $u$ uncorrelated in $x$. One then shows that $\Delta(u)$ does not change under renormalization \cite{LeDoussalWiese2011a,DobrinevskiLeDoussalWiese2011b}, i.e.\ the renormalized model is also given by  Eq.~(\ref{Delta-ren})
with $g=0$. This is the Brownian force model (BFM) introduced in \cite{LeDoussalWiese2011a}. It has a very simple {\em local} action, given by \Eq{8} with only the first term $\sim \sigma$. Since the velocity $\dot u_{xt}$ appears {\em linearly} in Eqs.~\eq{5}, \eq{6} and \eq{8}, 
the field theory is exactly solvable  \cite{LeDoussalWiese2011a,DobrinevskiLeDoussalWiese2011b},
\beq \label{11}
G[\lambda,f] = e^{ \int_{x,t} f_t \tilde u_{xt}^{\lambda}}\ .
\eeq
Here $\tilde u_{xt}^{\lambda}$ is the solution of the (exact) saddle-point or {\it instanton} equation  $\frac{\delta {\cal S}_{\lambda,f}}{\delta \dot u_{xt}}=0$, namely
\beq\label{10}
\eta\partial_t  \tilde u_{xt}^{\lambda}+(\nabla^2 -m^2 )\tilde  u_{xt}^{\lambda} + \sigma (\tilde u_{xt}^{\lambda})^2 = - \lambda_{xt}\ .
\eeq
The superscript $\lambda$ indicates that $\tilde u_{xt}^\lambda$ depends  on $\lambda_{xt}$.
This allows to calculate many observables exactly \cite{LeDoussalWiese2011a,DobrinevskiLeDoussalWiese2011b}. To simplify the calculations, one can 
express all observables
\begin{table}[h]
\newcommand{\n}{\hspace*{-1.2mm}}
\scalebox{1}{\begin{tabular}{|c|c||c|c||c|c|c|c|c|c|}
\hline 
  \n &  \n$d$\n & \n$\zeta$\n & \n$z$\n & \n$\tau$\n & \n$\tau_\phi$\n & \n$\alpha$\n & \n${\sf a}$\n & \n ${\sf a}_{\phi}$\n & \n$\gamma$  \n \\
\hline 
\hline
\n &  \n$1$\n & \n$1.25$\n & \n$1.433$\n & \n$1.11$\n & \n$0.4$\n & \n$1.17$\n & \n$-0.45$\n & \n $12.9$ \n & \n$1.57$  \n \\
\cline{2-10}
\n SR\n &  \n$2$\n & \n$0.75$\n & \n$1.56$\n & \n$1.27$\n & \n$-0.67$\n & \n$1.48$\n & \n$0.32$\n & \n $4.47$ \n & \n$1.76$  \n \\
\cline{2-10}
\n &  \n$3$\n & \n$0.35$\n & \n$1.75$ \n & \n$1.40$\n & \n$-3.71$\n & \n$1.77$\n & \n$0.75$\n & \n $3.43$ \n & \n$1.91$  \n \\
\hline
\n LR\n &  \n$1$\n & \n$0.39$\n & \n$0.77$ \n & \n$1.28$\n & \n$-0.56$\n & \n$1.51$\n & \n$0.39$\n & \n $4.63$ \n & \n$1.81$  \n \\
\hline
\end{tabular}}
\caption{Critical exponents obtained via the scaling relations using standard values
for $\zeta,z$ \cite{Numerics}. For the local avalanche exponents we consider a point, $d_\phi=0$.
}
\label{tab2}
\end{table}%
in units of $S_m=\sigma/m^4$ and $\tau_m=\eta/m^2$, equivalent to  
 setting $m^2=\sigma=\eta=1$. For a uniform source $\lambda_{xt}=\lambda$
one finds $\tilde u_{xt}^\lambda= \frac12 \left(1-\sqrt{1- 4 \lambda }\right)$ which
leads for a kick $f_{t}=m^2 w \delta(t)$ to $P_w(S)=\frac{w L^d}{2 \sqrt{\pi} S^{3/2}} e^{- (S-m^2 w)^2/4 S}$ and, in the limit of $w\to 0$, to 
the famous \cite{ABBM,Colaiori2008} mean-field size density 
$\rho(S) = \frac{L^d}{2 \sqrt{\pi} S^{\tau^{\rm MF}}} \rme^{-S/4}$ with $\tau^{\rm MF}=\frac{3}{2}$. 
Indeed, all observables containing only the center-of-mass are  equivalent  \cite{LeDoussalWiese2011a} to those of the phenomenological ABBM model \cite{ABBM,Colaiori2008}, which is nothing but the BFM in $d=0$. However, the BFM can go   further 
and allows to obtain the dependence on system size $L$, kick amplitude $w$, as well as {\em local} observables, such as 
the motion of a small piece of the interface, or the response to a local kick. 
Splitting $x=(x_\parallel,x_\perp)$ with $x_\parallel \in \mathbb{R}^{d_\phi}, x_\perp \in 
\mathbb{R}^{d-d_\phi}$, we focus on the submanifold $\phi$ of dimension $d_\phi$ given by $x=(x_\parallel,0)$. The {\it local size} of an avalanche on $\phi$ is
\beq \label{local}
S_\phi = \int \rmd t \int \rmd^{d_\phi} x_{ \parallel} \,\dot u_{ (x_\parallel,x_\perp=0),t} \ .
\eeq 
It is expressed  in units of $S_m^\phi=S_m m^{d-d_\phi}$, see below. Explicit solution
of \eqref{10} for the corresponding source $\lambda_{xt}=\lambda \delta^{d-d_\phi}( x_\perp)$ is
possible for $d_\phi=d-1$, leading to $\rho(S_\phi) = \frac{2 L^{d_\phi}}{\pi S_\phi} {\rm K}_{1/3}(2 S_\phi/\sqrt{3}) \sim S_\phi^{-\tau_\phi^{\rm MF}}$ 
in terms of a Bessel function \cite{LeDoussalWiese2008c}  , with (in that case) $\tau_{\phi}^{\rm MF}=\frac{4}{3}$.

Dynamical observables can be obtained from the solution $\tilde u^\lambda_{xt} = \frac{\lambda \theta(T-t)}{\lambda+(1-\lambda)\rme^{T-t}}$
of \eqref{10} with the source $\lambda_{xt} = \lambda \delta(T-t)$ \cite{LeDoussalWiese2011a,DobrinevskiLeDoussalWiese2011b}. Applying a kick at time $t=0$ and taking $\lambda \to -\infty$ selects $\dot u_{x,T}=0$, i.e.\ the avalanches of duration smaller than $T$. From $\tilde u^\lambda_{x,0}$
and using Eqs.~\eqref{9} and \eqref{11} one  obtains the PDF of durations as $P_w(T)=\frac{w L^d}{(2 \sinh T/2)^2} e^{-w L^d/(e^T-1)}$.
It converges to a Gumbel distribution for $w L^d \gg1$ (longest duration among many independent avalanches), while for $w L^d \ll1$ it leads to the known
mean-field duration-density \cite{Colaiori2008} with exponent $\alpha^{\rm MF}=2$. Calculating instead $\int_{t} \tilde u^\lambda_{xt} = - \ln(1-\lambda)$
for a constant driving $\dot w_t=v$, one obtains the stationary PDF of the {\it total instantaneous velocity},
\bea \label{tot} 
\dot {\sf u}_t := \int_{x} \dot u_{xt}\;,
\eea
as $P_v(\dot {\sf u}) = \dot {\sf u}^{-1+ v L^d} e^{-\dot {\sf u}}/\Gamma(v L^d)$,
in units of $v_m=S_m/\tau_m$. For $v L^d \ll 1$ it yields the density 
$\rho(\dot {\sf u}) = \frac{L^d}{\dot {\sf u}} \rme^{-\dot {\sf u}}$, in agreement with the $d=0$ velocity distribution 
 \cite{ABBM,Colaiori2008}.

\begin{figure}[t]
\fig{1}{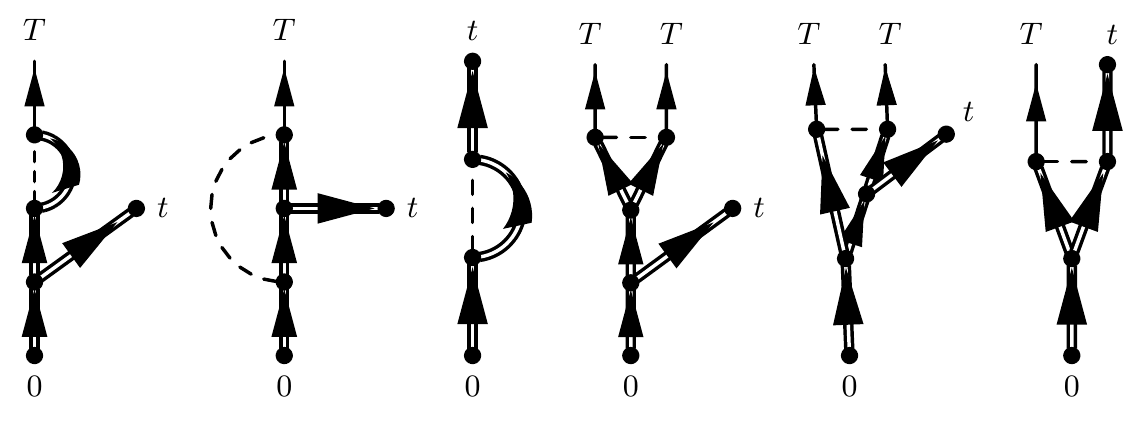}
\caption{Diagrammatic representation of the 1-loop corrections to the shape 
at fixed duration \eqref{MAIN} (similarly for \eqref{228}). Solid lines are response functions, doubled for
dressed ones, defined in \cite{LeDoussalWiese2011a}; they  account for the
non-vanishing expectation of $\tilde u_{xt}$ in  \Eq{10}. Dashed lines are  $g$-vertices, 
the other vertices are $\sigma$. 
Internal times and the loop momentum are integrated over.}
\label{fig:DiagsShape}
\end{figure}

\medskip

\paragraph{Field theory beyond the Brownian force model:}
It was shown in \cite{LeDoussalWiese2011a} that the BFM is the mean-field limit of
the field theory defined by Eqs.~\eq{G}, \eq{5}, i.e.\ it gives the joint multi-space-time-point
 velocity PDF in a single avalanche for $d \ge d_{\rm c}$ \footnote{with suitably renormalized values for $\sigma$ and
$\eta$, including corrections in $\ln(1/m)$ for $d=d_{\rm c}$ see \cite{LeDoussalWiese2011a}.}.
Moreover, including 
the term $\sim g$ in \Eq{Delta-ren} is {\em sufficient} to obtain {\em the complete 1-loop corrections}, i.e.
to calculate these distributions to first order in an expansion in $\epsilon =d_{\rm c}-d$,
with $g={\cal O}(\epsilon)$ at the fixed point. The velocity density was obtained to one loop \cite{LeDoussalWiese2011a}, with 
a non-trivial tail for $\dot {\sf u} \gg 1$, and a power-law singularity\footnote{%
\label{footnote1}It holds for depinning of an interface, all ${\cal O}(\varepsilon)$ results here can be extended
to a {\it periodic} object in $d=d_{\rm c}-\varepsilon$ by the replacement $\varepsilon \to \frac{3}{2} \varepsilon$ in all
formulas; generally $\varepsilon \to \frac{3}{2} (\varepsilon-\zeta)$.
}\beq \label{veldistrib} 
\rho(\dot {\sf u}) \simeq_{\dot {\sf u} \ll 1} \frac{C L^d}{\dot {\sf u}^{\sf a}}, \quad {\sf a} = 1-\frac29 \epsilon + O(\epsilon^2)\,,
\eeq
with $C=1-\frac{\epsilon}{9} (4 \gamma_{\rm E} + \frac{1}{2} - 2 \ln 2)$.

\medskip
\paragraph{Exponent relations:} 
At the level of the field theory of depinning, i.e.~\Eq{5} to two loops, and for
avalanches \Eq{8} to one loop, until now we have found only two independent 
renormalizations, one for the disorder\footnote{Since the whole function
$\Delta(u)$ is relevant for $d < d_c$, in principle one needs 
an infinity of renormalizations \cite{DSFisher1986,ChauveLeDoussalWiese2000a},
however those are not independent at the fixed point.} 
 $\sigma \to \sigma_m$, and one for 
the friction $\eta \to \eta_m$, leading to two independent scales in any dimension $d$:
\beq \label{units}
S_m=\sigma_m/m^4 \sim m^{-(d+\zeta)} \ , \quad \tau_m =\eta_m/m^2 \sim m^{-z} \ .
\eeq 
This suggests that avalanche exponents, 
such as $\tau$, $\alpha$ and ${\sf a}$ {\em are not independent}, but instead related to the  
roughness  $\zeta$ and dynamical exponent $z$. Starting with 
the Narayan Fisher (NF) conjecture \cite{NarayanDSFisher1992b} for $\tau$, this has
been a recurrent question in the field \cite{ZapperiCizeauDurinStanley1998},
and, for the velocity exponent ${\sf a}$, an outstanding  one.

We now reexamine and extend the NF conjecture using dimensional and field
theoretic arguments. Restoring units (i.e.\ all factors of $m$), the size density (per unit $w$)
takes the form $\rho(S) = L^d S_m^{-2} (S_m/S)^{-\tau} f(S/S_m)$ 
with $f(0)$ a finite constant. The NF conjecture is equivalent to stating that the size density {\it per unit force}, $\rho_f(S)=m^{-2} \rho(S)$, 
has a finite (infrared-cutoff independent) limit $m \to 0$, i.e.
\beq\label{rhof}
\rho_f(S) \sim L^d {S^{-\tau}} f(S/S_m)\ , 
\eeq
up to a constant prefactor. This implies $S_m^{2-\tau} \sim m^2$, i.e.\ 
\beq\label{14}
\tau = 2- \frac2{d+\zeta}\ .
\eeq
In the field theory, one can use the exact relation\footnote{$\langle ... \rangle_{\lambda}$ denotes 
averages w.r.t. the action ${\cal S}_{\lambda,f=0^+}$ in Eq.~\eq{5}.}
\beq \label{resp1} 
\int \rmd S (\rme^{\lambda S}-1 ) \rho_f(S)= L^d \langle \tilde u_{x,t=0} \rangle_{\lambda} \ .
\eeq
Upon the same assumption (\ref{rhof}) the result (\ref{14}) can equivalently be obtained  from Eq.~(\ref{resp1}):
In the action \eq{5} the term $\int_{xt} \tilde u_{xt} m^2 \dot u_{xt}$ is protected by the
statistical tilt symmetry, hence the response field has dimension $\tilde u_{xt}\sim m^{d-2+\zeta}$. 
Matching the l.h.s.\ at $- \lambda=1/S_m$ yields $(-\lambda)^{\tau-1}\sim S_m^{1-\tau} \sim m^{d-2+\zeta}$
recovering\footnote{The r.h.s.~takes the form, $\langle \tilde u_{xt=0} \rangle_{\lambda} = m^{d-2+\zeta} g(\lambda m^{-(d+\zeta)})$.
For $1<\tau<2$ it has a finite $m\to 0$ limit $\sim (-\lambda)^{\tau-1}$. 
} 
\Eq{14}. 
 The field theory confirms that the quantity which must have a $m \to 0$ limit is 
$\rho_f(S)$, and not $\rho(S)$, in order that (\ref{14}) holds.

Consider now the distribution of the total velocity $\dot {\sf u}$ defined in \Eq{tot}, and define  the 
density {\it per unit force change} $\dot f=m^2 v$,
\beq
\rho_f(\dot {\sf u}) = \partial_{\dot f} P_v(\dot {\sf u})|_{\dot f=0^+} \ .
\eeq 
It diverges as $m^2 \rho_f(\dot {\sf u})  \sim \frac{L^d}{(v_m)^2} (v_m/\dot {\sf u})^{\sf a}$ by dimensional analysis,
where $v_m = S_m/\tau_m$. The existence of a massless limit for $\rho_f(\dot {\sf u})$ implies $(v_m)^{{\sf a}-2} \sim m^2$; 
hence, from Eq.~(\ref{units}) we obtain the new relation
\bea \label{NFa} 
{\sf a} = 2- \frac2{d+\zeta-z}\ .
\eea 
In the field theory,  this identity can be derived from 
\beq \label{rel3}
\int \rmd\dot {\sf u} ( e^{\lambda \dot {\sf u}}  - 1)  \rho_f(\dot {\sf u}) = L^d \int_t \langle \tilde u_{xt} \rangle_{\lambda}
\eeq     
with the source $\lambda_{xt}=\lambda \delta(t)$ and
an additional integral over the time where the avalanche was triggered.
Assuming that a massless limit exists for $\rho_f(\dot {\sf u})\sim \dot {\sf u}^{-{\sf a}}$, we can match the
l.h.s of Eq.~(\ref{rel3}) at $\dot {\sf u}  \sim v_m$, as $(-\lambda)^{{\sf a}-1}\sim v_m^{1-{\sf a}}$ 
and identify its mass dimension as $\sim m^{d-2+\zeta-z}$ from the r.h.s., leading
again to  Eq.~(\ref{NFa}).

This can be generalized to local avalanche observables. Assuming 
again a massless limit for densities per unit force one finds $(S^\phi_m)^{\tau_\phi-2} \sim m^2$
and the local avalanche-size density
\beq\label{18}
\rho(S_\phi) \sim_{S_\phi\ll S^\phi_m}  {S_\phi^{-\tau_\phi}}\ ,\qquad \tau_\phi = 2- \frac2{d_\phi+\zeta}\ .
\eeq
For the local velocity density one finds $(v^\phi_m)^{{\sf a}_\phi -2} \sim m^2$ 
where $v^\phi_m = S^\phi_m/\tau_m$ is the natural unit, and
consequently 
\beq\label{18bis}
\rho(\dot {\sf u}_\phi) \sim_{\dot {\sf u}_\phi \ll v^\phi_m}   \dot {\sf u}_\phi^{{-\sf a}_\phi}\ ,\qquad {\sf a}_\phi = 2- \frac2{d_\phi+\zeta-z}\ .
\eeq
Similar arguments for the duration distribution lead to
\beq\label{e:alpha}
\rho(T) \sim_{T\ll \tau_m}  {T^{-\alpha}}\ ,\qquad \alpha = 1+ \frac{d-2+\zeta}{z}\ ,
\eeq
recovering the result of \cite{ZapperiCizeauDurinStanley1998}
obtained by simple scaling from (\ref{units}) and the variable change  $\rmd S\, \rho(S) = \rmd T\, \rho(T)$. 
The mean avalanche size at fixed duration is likewise given by
\beq\label{19}
\left<S \right>_T \sim_{T\ll \tau_m} T^{\gamma}\ , \qquad \gamma = \frac{d+\zeta}{z}\ .
\eeq
For LR-elasticity $q^2 \to |q|$ (in Fourier) the predictions change as indicated
on table \ref{tab1}, where all results are summarized. (The formula for
$\gamma$ remains the same). 


In summary these scaling relations should hold, provided only two
independent renormalizations are sufficient to render the field theory
of depinning finite. The fact that $\dot f_{xt}$ and $ \lambda_{xt}$ are  linear perturbations
of the depinning action suggests that they cannot induce other renormalizations. Numerical values predicted by these conjectures
are indicated on table \ref{tab2}; it is important to check them in numerics and
experiments\footnote{\label{footnote2}Their validity may not extend to all cases: (i) in $d=0$ for SR disorder the NF conjecture fails since $\tau=0$,
$\zeta=2$ (plus logarithms) \cite{LeDoussalWiese2008a}, a case dominated by extreme value statistics
(ii) (\ref{resp1}),(\ref{rel3}) are ultraviolet divergent for exponents $>2$.
}.

\medskip

\paragraph{The shape at fixed duration:}
The shape of an avalanche conditioned on its duration $T$ is obtained from our field theory in
an expansion in $\varepsilon=d_{\rm c}-d$. The calculation is involved, 
and we only sketch its diagrammatic representation  in Fig. \ref{fig:DiagsShape}. The general result is lengthy,
hence we   only display its universal\footnote{%
In Eqs.~\eqref{veldistrib} and \eqref{MAIN}, $T$ and $\dot {\sf u}$ are in
units of $\tau_m,v_m$. {Restoring units and using (\ref{units}) and (\ref{19}) all factors of $m$ cancel in Eq. \eqref{MAIN}.}
}  limit for short duration $T \ll \tau_m$, 
\begin{eqnarray}\label{MAIN}
\left< \dot {\sf u}\left(t = xT \right) \right>_T &=& 2 {\cal N}\Big[ Tx(1-x)\Big]^{\gamma-1} \\
&& \times \exp\bigg(- \frac{16 \varepsilon}{9 d_{\rm c}} \bigg[ \text{Li}_2(1-x)-\text{Li}_2\Big(\frac{1-x}{2}\Big)\nn\\
&& ~~~~~~+\frac{x \log (2 x)}{x-1}+\frac{(x+1) \log
   (x+1)}{2(1- x)}\bigg] \bigg), \nn
\end{eqnarray}
%
%
with $d_{\rm c}=4$ for SR and $d_{\rm c}=2$ for LR elasticity. 
The  scaling $\sim T^{\gamma-1}$ is 
expected from the sum rule
$\int_0^T \rmd t\,  \left< \dot {\sf u}(t) \right>_T=\left<S\right>_T \sim T^{\gamma}$
and our calculated value $\gamma=2 - \frac{4} {9 d_{\rm c}} \varepsilon$ is consistent to ${\cal O}(\epsilon)$ with \Eq{19} \footnote{using $\zeta=\frac{\varepsilon}{3}$ and $z=2-\frac{2 \varepsilon}{9}$ to one loop \cite{NattermannStepanowTangLeschhorn1992,ChauveLeDoussalWiese2000a}.}. The exponential factor in (\ref{MAIN}) is regular at $x=0$ and $x=1$. The
   singular part of the shape, $[x(1-x)]^{\gamma-1}$, is thus symmetric, as 
   anti\-cipated on phenomenological grounds \cite{LaursonIllaSantucciTallakstadyAlava2013}, and derived here from first principles. 
We chose to display \Eq{MAIN} in an exponentiated form so that the
amplitudes, ${\cal N}_{{\rm SR}} =e^{- \frac{\varepsilon}{9} [ \gamma_{\rm E}
   -1-2 (\ln 2)^2-\frac{\pi ^2}{3}]}$, ${\cal N}_{{\rm LR}} = e^{- \frac{2 \varepsilon}{9} [ \gamma_{\rm E}
   -2-2 (\ln 2)^2 -\frac{\pi ^2}{3}]}$ cancel if one plots the
   {\em normalized shape}    
\begin{figure}[t]
\Fig{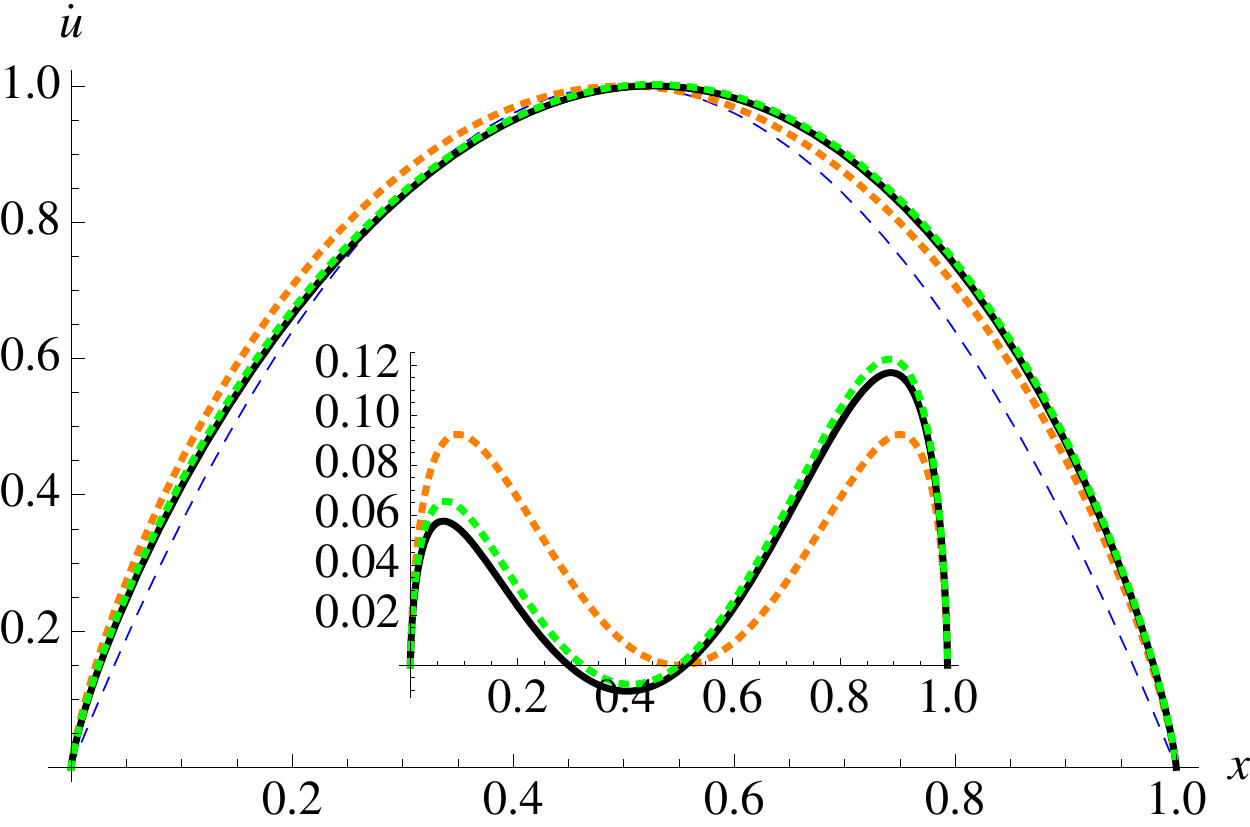}
\caption{(Universal) normalized shape of an avalanche (of short duration $T \ll \tau_m$), for an interface ($d=2$) with SR elasticity.
Plotted is the total velocity $\dot {\sf u}(t)$ at time $t = x T$ from \Eq{MAIN}, normalized to unit maximum (black thick solid line), compared to: (i) the MF shape $\sim x(1-x)$ (blue, dashed, thin line); (ii) a symmetric scaling-ansatz $\dot {\sf u} \sim [ Tx(1-x)]^{1- \frac{\epsilon}{9}}$ (orange, dot-dashed, thick); (iii) the
approximation \eq{21} (green dots),  close to the exact result. Inset: {\em ibid.} with the MF shape subtracted.}
\label{f:shapeSRd=2}
\end{figure}%
as in Fig.~\ref{f:shapeSRd=2}. The result (\ref{MAIN}) is exact${}^{\ref{footnote1}}$ 
up to terms of order ${\cal O}(\varepsilon^2)$. 
Note that, at variance with mean field ($\varepsilon=0$), {\it the full shape is not symmetric} under  $x\to 1-x$.
In fact, the complicated factor in the exponential in (\ref{MAIN}) turns out to be almost
linear, hence a good approximation (ignoring constant prefactors) is
\beq\label{21}
\left< \dot {\sf u}\left(t = xT \right) \right>_T  \sim \Big[ Tx(1-x)\Big]^{\gamma-1} \exp\Big( {\cal A}_d (\textstyle \frac12-x) \Big) .
\eeq 
The asymmetry ${\cal A}_d$ is defined, e.g. as the slope at $x=\frac{1}{2}$ of the exponential in \Eq{MAIN}.
Close to $d=d_{\rm c}$ we obtain
\beq \label{eq21}
{\cal A}_d  \approx \textstyle - 0.336 \left(1-\frac d {d_{\rm c}}\right) \ .
\eeq 
An extrapolation of \Eq{MAIN} to  $d=2$ for SR elasticity,  and  $d=1$ for LR elasticity, 
is plotted in Fig.~\ref{f:shapeSRd=2}. 

Hence we find a {\it negative asymmetry} near the upper critical dimension. This is 
consistent with numerical simulations for
SR elasticity in dimensions $d=2,3$, which suggest that 
avalanches are skewed towards the end, i.e.\ with \Eq{21} for ${\cal A}_{d=2} \approx -0.065 \pm 0.01$ \cite{LaursonPrivate}.
On the other hand, numerical results in $d=1$ for both SR and LR elasticity suggest  skewing 
towards the beginning \cite{LaursonIllaSantucciTallakstadyAlava2013} 
with positive asymmetries ${\cal A}_{d=1}\approx 0.08 \pm 0.02$ (SR) and ${\cal A}_{d=1} \approx 0.02 \pm 0.02$ (LR).
To check whether this sign change could be accounted for at 1-loop order,  we performed 
a fixed-$d$, weak-disorder expansion; it does not seem to predict
this effect \cite{LeDoussalWieseToBePublished}. Hence this sign change, if confirmed, would be a higher-loop effect\footnote{Other differences, such as in the
roughness exponents between equilibrium and driven dynamics are also due to two loops
\cite{ChauveLeDoussalWiese2000a}.}. 
Note that the approximate time-reversal symmetry is hard to explain intuitively
since ``active" regions within an avalanche split  over time and become disjoint in space
(see e.g.\ Fig.\ 1 in  \cite{LaursonIllaSantucciTallakstadyAlava2013}). Nevertheless, the ensemble-averaged 
velocity is {\em almost} time-reversal symmetric. The small asymmetries 
 thus result from a delicate balance of several $d$-dependent effects\footnote{Note that non-zero wave-vector observables exhibit a positive asymmetry even within mean-field theory
\cite{LeDoussalWiese2011a}.}. It would  be important to thoroughly test our predictions in $d=2,3$. 


\begin{figure}[t]
\Fig{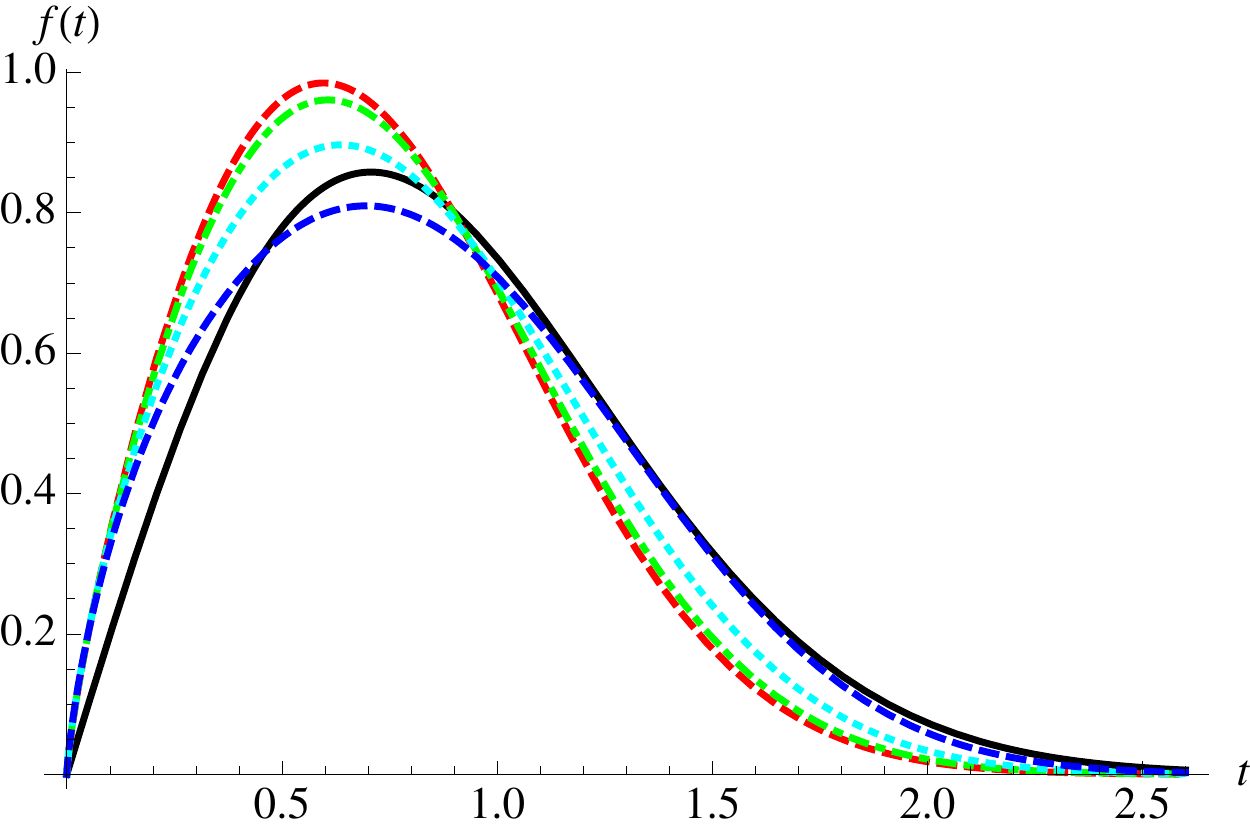}
\caption{The shape at fixed size, as given by Eq.\ (\ref{239}). Mean field (black
solid line). The remaining curves are for  $\varepsilon=2$: small $S/S_m=0^+$ limit (red dashed)
and $S/S_m=1$,
$10$, $30$ (green dot-dashed, cyan dotted,
and blue dashed). }
\label{f3c}
\end{figure}

\paragraph{The shape at fixed size:}
We propose to measure a new observable,  depending only minimally on  the criterion  to define the end of an avalanche. It is the mean velocity as a function 
of time, given that the avalanche size is $S$. Scaling suggests
that 
\bea \label{shapeS} 
\langle \dot {\sf u}(t) \rangle_S = \frac{S}{\tau_m} \Big(\frac{S}{S_m}\Big)^{\!-\frac{1}{\gamma}} f\bigg( \frac{t}{\tau_m}\Big(\frac{S_{m}}{S}\Big)^{\!\frac{1}{\gamma}} \bigg)
\eea 
with $\int_0^{\infty} \rmd t\, f(t) = 1$, where $f(t)$ may depend on $S/S_m$. In mean field, the scaling function $f(t)$
is independent of $S/S_m$ \cite{DobrinevskiLeDoussalWiese2013},
and reads\bea
f_0(t) = 2 t e^{- t^2}  \quad , \quad \gamma=2\ .
\eea 
To one loop, i.e.\ ${\cal O}(\varepsilon)$, for SR elasticity, we obtain
\bea
f(t) = f_0(t) - \frac{\epsilon}{9} \delta f(t)  \quad , \quad \gamma=2 - \frac{\epsilon}{9}\ ,
\eea 
consistent with (\ref{19}).
Expressions for any $S/S_m$ are lengthy and we display
only the universal small-avalanche limit:
\bea
\label{228}
 \delta f(t) &=&  \frac{f_0(t)}{4} \bigg[\pi  \left(2
   t^2+1\right) \text{erfi}(t)   +2 \gamma_{\rm E} 
   \left(1-t^2\right)-4
   \nn\\&&~~~~~~~~-2 t^2 \left(2 t^2+1\right) \,
   _2F_2\left(1,1;\frac{3}{2},2;t^2\right)\nn\\&& 
   ~~~~~~~~-2 e^{t^2}
   \Big(\sqrt{\pi } t\,
   \text{erfc}(t)-\text{Ei}\left(-t^2\right)\Big)\bigg]\ . ~~~~~~~
\eea 
It satisfies $\int_0^{\infty} \rmd t\, \delta f(t) =0$. The asymptotic behaviors  are \begin{align} \label{asympt} 
f(t) &\simeq_{t \to 0} 2 A t^{\gamma-1} \\
 f(t) &\simeq_{t \to \infty}  2 A' t^{\beta} e^{- C t^\delta} \label{asympt2},\quad \textstyle  \delta = 2 + \frac{\epsilon}{9} , \quad \beta = 1- \frac{\epsilon}{18},  
\end{align}
with $A' = 1+ \frac{\epsilon}{36}  (5 - 3 \gamma_{\rm E} - \ln 4)$ and 
$C = 1+ \frac{\epsilon}{9} \ln 2$. The amplitude $A=1 + \frac{\epsilon}{9} (1-\gamma_{\rm E})$
leads to the same universal short-time behavior as in \eqref{MAIN}, near the avalanche beginning $t\ll T$. Extrapolation for the function $f(t)$ in $d=2$ is plotted in Fig \ref{f3c}.
We use
\beq \label{239}
f(t) \approx 2 t e^{-C t^\delta} B \exp\!\left(- \frac{\epsilon}{9} \!\left[\frac{\delta f(t)}{f_0(t)}{-}t^2 \ln (2t)\right]  \right),
\eeq
with $B$ chosen s.t.\ $\int_0^{\infty} \rmd t f(t) = 1$.  \Eq{239} is exact  to ${\cal O}(\varepsilon)$
and obeys (\ref{asympt}),  (\ref{asympt2}).
As one sees in Fig.~\ref{f3c},  all avalanches start similarly, while for larger (scaled) sizes they flatten out 
and extend to longer times.

In conclusion, based on the FRG field theory of disordered elastic interfaces, we have derived  new avalanche scaling relations, 
and calculated the shape of an avalanche, both at fixed duration and at
fixed size. We hope our predictions  stimulate new experiments and simulations. 

\acknowledgments We thank L.~Laurson for sharing his unpublished data, as well as A.~Kolton,
G.~Durin, S.~Santucci and A.~Rosso for stimulating discussions. 
This work was supported by PSL grant ANR-10-IDEX-0001-02-PSL. 


\end{document}